\documentclass[referee]{aa}
\usepackage{graphicx}
\begin{document}
\title{
Superfluid vortices in neutron stars} 
\author{ \O ystein Elgar\o y \inst{1}
\thanks{\emph{Present address:} Institute of Astronomy, 
University of Cambridge, 
Madingley Road, Cambridge CB3 0HA, UK}
 \and Fabio Vittorio De Blasio \inst{2} }  
\institute{Institute of Theoretical Astrophysics, 
University of Oslo, Box 1029,
N - 0315 Oslo, Norway 
\and 
Department of Physics, University of Oslo, P.O. Box 1048 Blindern 
N - 0316 Oslo, Norway 
}
\date{}
\authorrunning{Elgar\o y and De Blasio}
\abstract{
A microscopic, quantum mechanical model for neutron vortices in 
the crust of a neutron star is presented.  
After a brief introduction to the Bogoliubov-de Gennes equations,  
which form the basis for our calculations, we present    
results for density distributions, vortex core sizes and vortex energies,  
both for an isolated neutron vortex and for the case 
when the vortex core overlaps with a cylindrical nucleus.  
Earlier results on the vortex core size are confirmed, 
indicating a much less dramatic variation of the 
vortex core size with density than predicted by the BCS formula.   
}
\maketitle

\section{Introduction}

A natural probe of the internal stucture 
and dynamics of neutron star crusts 
 is   represented by pulsar glitches, which are   sudden 
accelerations of the star's rotational frequency. 
The observation  of more than seventy events in 
about thirty pulsars  provides
 a general view of the phenomenon.
At present the standard model to explain pulsar glitches  
focuses on 
  the interaction of 
superfluid vortices with the crust of the neutron star 
(\cite{alpar_pines}). Some variants 
include   crust breaking, resulting in a      
starquake, or  stellar plate tectonics due to the action  
of  magnetic threading of proton vortex lines.  

The standard model  is built in  analogy to  
type II superconductors, where the attractive interaction of vortex cores 
with the lattice defects of the metal prevents the motion of vortex lines 
through a metal sample.  
An important but unknown parameter  is  
the pinning energy of a vortex to a nucleus,   the difference in   
energy between the configurations in which a vortex and a 
nucleus are superimposed and the one in which they are far apart and 
non-interacting. 
Usually this quantity is calculated simply as the loss of 
condensation energy between the two configurations. Since the 
core of a vortex is composed of normal matter and the neutron superfluid 
which occupies the nuclear volume is, for sufficiently 
high densities, a weaker superfluid (i.e. has a 
smaller local pairing gap) than the superfluid outside the nuclei, 
it is energetically favourable for a vortex to remain attached to 
a nucleus. However, this analysis neglects important effects like the 
change in kinetic energy 
(Epstein \& Baym 1988; Pizzochero, Viverit \& Broglia 1997) 
and the presence of bound states (\cite{deblasio99}). 

Furthermore, the vortex is depicted as a 
cylinder composed of uniform normal matter with a sharp radius 
equal to the BCS coherence length of the superfluid 
(the BCS coherence length is the mean square 
radius of Cooper pairs, see \cite{noi_cohere} for the case 
of hadronic matter). This 
is a very rough model since  
 the gap changes smoothly inside the vortex and 
not abruptly, as assumed. 
 The vortex-nucleus  
interaction results from 
effects requiring a full quantum mechanical  
solution of the BCS equations in the cell. 
Another approach which has been attempted is  the 
Ginzburg-Landau  model (\cite{baym}). This 
model  provides a good treatment of the kinetic energy density, which 
is an important  ingredient in the energy balance 
within the interacting volume.    
 Unfortunately, the pairing is not well 
treated in the Ginzburg-Landau equations   since the two main 
conditions for their applicability, that the temperature should be 
close to the critical one and that there should be only slow and 
smooth spatial variations in the order parameter, are not fulfilled 
in a neutron star crust.  

The relevant quantity, the vortex-nucleus 
pinning force, can in principle  
be   calculated 
from the variation of  the pinning energy with respect to the 
vortex-nucleus distance.  An approximate 
expression can be found by  dividing the 
pinning energy by the minimal distance where the two objects 
can be considered as separate.     
The pinning force enters in two important 
ways in the standard model for pulsar glitches. 
Vortices are released from the crust 
when   the frequency lag between the superfluid 
and the crust reaches  
a critical value. They   unpin in clusters,   
decreasing suddenly     
 the   angular 
momentum of the superfluid. The glitch results from   
 the conservation of angular momentum,  causing    
 the crust to increase its spinning rate.  
Therefore, the critical frequency is proportional to the value of the 
local pinning force.   

The second issue where pinning force is relevant  
is     postglitch relaxation. This is particularly interesting, because 
time analysis following a glitch 
is now available for some glitches and  especially for the 
Vela pulsar. 
In the 
vortex creep model, vortices that have abandoned their 
pinning centers repin with other nuclei  within  
characteristic times which  also depend on the 
value of the pinning force.  
The pinning force 
is of the order $ E_p/\xi $, where $\xi$ is an appropriate length scale 
of the vortex--nucleus interaction potential, roughly the 
size of the vortex core, and $ E_p $ is 
the pinning energy. This length is usually taken  
to be the BCS coherence length, but recent calculations 
(\cite{deblasio99})  show that the 
vortex core can be significantly smaller than the BCS coherence 
length. This indicates that if the 
pinning energies were comparable, the pinning force should  
be larger.    
It is clear that a better understanding of 
the pinning model for pulsar glitches will be possible 
when more reliable 
calculations of the structure of a superfluid neutron vortex 
are available.

Again, the study of superfluid vortices in neutron stars is 
not exhausted by considering solely 
neutron vortices in the crust. 
 Below the inner crust, 
neutrons couple in the $^3P_2$ channel to generate a 
triplet superfluid, while protons, having a much smaller density, 
couple in the singlet channel. The flux of protons around the 
vortex core generates a magnetic field parallel to the 
rotational axis.  
In addition, a large number ($10^{19}$ cm$^{-3}$)  
of proton flux lines is generated by the very 
high magnetic field ($\sim 10^{12} $G) inside the core.  
The interaction 
of these flux lines with neutron vortices drifting towards the 
exterior of the star might be 
an efficient mechanism for magnetic field decay (see for example  
\cite{ruderman}). 

The study  of these physical situations 
may require  a more accurate model for vortices in hadronic matter.  
A fully quantum mechanical model of vortices  
in type II superconductors has been extensively investigated 
in some publications (\cite{gygi,jap}).     
It is possible to study the 
properties of neutron vortices parallelling the model for 
superconductors.   
Experiments with scanning-tunnelling microscopy 
and  refined numerical  
calculations (\cite{gygi}) have  
confirmed the theoretical prediction \cite{caroli,bardeen}  that  
bound states are  formed in the center of the core of a vortex 
in type-II superconductors. 
The pairing gap  decreases to zero at the 
center of the core  and grows to an asymptotic value 
within distances of the order of or larger than the 
coherence length. 
 
In the present paper we extend our previous work (\cite{deblasio99}), 
hereafter Paper I,  
on the microscopic structure of neutron vortex lines. 
In Paper I we made use of  
the Bogoliubov-de Gennes  
equations (\cite{degennes,ketterson}), that have been successfully 
developed  in  
studies of vortices in type II superconductors (\cite{gygi,jap}) and 
more in general non-homogeneous superconductivity. 

In Paper I 
we were mostly interested 
in the vortex core size, since this quantity influences the 
pinning energy and force.  We will here  
continue our study of this quantity, providing more complete results  
and extending our calculations towards a fully microscopic calculation 
of pinning energies.  Furthermore, we will show results for quantities 
such as  density distributions, pairing potentials and vortex tension.  
The calculations presented here are also more refined than those in 
Paper I, as they 
are fully three-dimensional and include effects like the 
Hartree mean field and a density-dependent pairing force.   
Furthermore, since our earlier paper was brief on the technicalities, 
we will give a more detailed account of the formalism and 
the numerical solution of the relevant equations.  

\section{The model} 

The $^1S_0$ neutron superfluid in the inner crust is spatially 
nonhomogeneous due to the presence of a nuclear lattice and the 
vortices induced in the superfluid by the  rotation of the star.  
To describe this system one needs to go beyond the standard 
BCS formalism for fermion pairing.  A formalism for this has 
existed in solid state physics for several years in the form of 
the Bogoliubov-de Gennes (BdG) equations (\cite{degennes}).  
For readers not familiar with these, we will give a  
sketch of their derivation.  

The theory is most conveniently formulated in terms of the 
neutron field operators $\psi_\sigma({\bf r})$ and $\psi_\sigma ^\dagger 
({\bf r})$ which respectively destroy and create a neutron with 
spin projection $\sigma$ in position ${\bf r}$.  
As a starting point, we choose a pure pairing Hamiltonian 
\begin{equation}
H=H_0+H_{{\rm int}}=H_0-|g| \int d^3 r   
\psi_{\uparrow}^\dagger ({\bf r}) 
\psi_{\downarrow}^\dagger ({\bf r}) \psi_\downarrow ({\bf r}) 
\psi_\uparrow ({\bf r}), \label{eq:eq369} 
\end{equation}
where $|g|$ is the pairing strength (we use the notation $|g|$ to 
avoid confusion about the sign of the pairing strength, which in 
some works is defined to be negative, in others to be positive).  
This corresponds to a zero-range 
pairing interaction of the form 
\begin{equation} 
v({\bf r},{\bf r'})=-|g| \delta ({\bf r}-{\bf r'}).  
\end{equation} 
 and necessitates the use of a momentum 
cutoff confining the interaction to a narrow set of states 
near the Fermi surface.  
Note that in general the pairing strength is space dependent 
due to density variations induced by the mean field or by the 
vortex.  We will return to this point later, for the time being 
we assume that it is a constant.  

The one-body hamiltonian  has the form 
\begin{equation} 
H_0=\int d^3 r [\psi_{\uparrow}^\dagger ({\bf r})  
h_0 
 \psi_{\uparrow} ({\bf r})+ 
  \psi_{\downarrow}^\dagger ({\bf r})  
h_0 
 \psi_{\downarrow} ({\bf r})] 
\label{eq:spham}  
\end{equation} 
where   
\begin{equation} 
h_0=\hat{t}+W({\bf r})+S({\bf r})=\hat{t}+Y({\bf r})
\end{equation}
and 
\begin{equation}
\hat{t}=-\frac{\hbar^2 \nabla ^2}{2m} -\lambda.  
\label{eq:eq382}
\end{equation}
In the above equations $\lambda$ is the chemical potential,      
$W({\bf r})$ is an external potential representing a 
nucleus, $S({\bf r})$ is a self-consistent 
potential generated by the neutrons themselves while $ Y({\bf r})$ stands 
for the total field. 

By selective averaging, or, more formally, through applying 
the saddle-point approximation to the grand canonical partition 
function which follows from this Hamiltonian, one can 
derive a mean field Hamiltonian 
\begin{equation}
H_{{\rm MFA}}=H_0+\int d^3 r [\psi_\uparrow^\dagger ({\bf r}) 
\psi_\downarrow ^\dagger ({\bf r}) \Delta ({\bf r}) 
+\Delta ^* ({\bf r}) \psi_\downarrow ({\bf r}) \psi_\uparrow 
({\bf r})]+\zeta  
\label{eq:eq370}
\end{equation}
with 
\begin{equation}
\Delta({\bf r})=|g|\langle \psi_\downarrow ({\bf r}) 
\psi_\uparrow ({\bf r}) \rangle 
\label{eq:eq371}
\end{equation}
being sometimes called ``the complex order parameter'', 
``the pairing potential'', or ``the pair function''. 
Since this is a mean-field approximation, quantum fluctuations in the 
order parameter is neglected.  
The quantity 
\begin{equation}
\zeta=\int d^3 r {|\Delta({\bf r})|^2\over |g|}
\end{equation} 
is a c-number which plays no role in the free energy minimization 
leading to the Bogoliubov-de Gennes equations, but which is important 
in the energy calculations.
Here, and in the following $\langle \ldots \rangle$ denotes 
thermal averages.  
The mean field Hamiltonian (\ref{eq:eq370}) can be diagonalized by 
a canonical Bogoliubov transformation 
\begin{eqnarray}
\alpha_{i,\uparrow}(t)&=&\int d^3 r [U_i^*({\bf r})\psi_\uparrow
({\bf r},t)-V_i^* ({\bf r}) \psi_\downarrow ^\dagger ({\bf r},t)] 
\label{eq:eq372} \\
\alpha_{i,\downarrow}^\dagger(t) &=& \int d^3 r [U_i ({\bf r}) 
\psi_\downarrow ^\dagger ({\bf r},t) + V_i({\bf r})\psi_\uparrow 
({\bf r},t)], \label{eq:eq373} 
\end{eqnarray}
with the inverse transformation being 
\begin{eqnarray}
\psi_\uparrow ({\bf r},t)&=&\sum_i [U_i({\bf r})\alpha_{i,\uparrow}(t)
+V_i^* ({\bf r})\alpha_{i,\downarrow}^\dagger (t)] \label{eq:eq374} \\ 
\psi_\downarrow ^\dagger ({\bf r},t)&=& \sum_i [U_i^*({\bf r}) 
\alpha_{i,\downarrow}^\dagger (t) -V_i({\bf r})\alpha_{i,\uparrow}(t)]. 
\label{eq:eq375}
\end{eqnarray}
In these equations, $i$ is a label for the single-particle eigenstates. 
The coefficients $U$ and $V$ of this Bogoliubov transformation must 
obey the canonical relations 
\begin{eqnarray}
\sum_i [U_i ({\bf r}) U_i^* ({\bf r'})+V_i({\bf r'})V_i^*({\bf r})]
&=&\delta({\bf r}-{\bf r'}) \label{eq:eq376} \\ 
\sum_i [ U_i ({\bf r})V_i^*({\bf r'})-U_i({\bf r'})V_i^*({\bf r})]
&=&0 \label{eq:eq377} \\ 
\int d^3 r [U_i({\bf r})U_{i'}^{*}({\bf r})+V_i({\bf r})V_{i'}^*({\bf r})] 
&=&\delta_{ii'} \label{eq:eq378} \\ 
\int d^3 r [U_i({\bf r}) V_{i'}({\bf r})-U_{i'}({\bf r})V_i({\bf r})] 
&=&0, \label{eq:eq379} 
\end{eqnarray}
in order for the quasiparticle operators to follow Fermi statistics. 
Carrying out this transformation, the Hamiltonian takes the 
simple form 
\begin{equation}
H_B=U_0+\sum_i E_i(\alpha_{i,\uparrow}^\dagger \alpha_{i,\uparrow} 
+\alpha_{i,\downarrow}^\dagger \alpha_{i,\downarrow}).  
\label{eq:eq380}
\end{equation}
The constant term $U_0$ is the ground state energy of the 
system.  The Bogoliubov-de Gennes equations for the transformation 
coefficients which bring the Hamiltonian to the diagonal form $H_B$ 
follow from the Heisenberg equations of motion for the field 
operators, $i\stackrel{.}{\psi}=[\psi,H_B]$, that is, in matrix form  
\begin{equation}
i\frac{\partial}{\partial t} \left( \begin{array}{c} 
\psi_\uparrow ({\bf r},t) \\ \psi_\downarrow ({\bf r},t) 
\end{array} \right)=\left( \begin{array}{cc} \hat{t}+Y({\bf r}) 
& -\Delta({\bf r}) \\ -\Delta^*({\bf r}) & -\hat{t}-Y({\bf r}) 
\end{array} \right) \left( \begin{array}{c}\psi_\uparrow ({\bf r},t) 
\\ \psi_\downarrow ({\bf r},t) \end{array} \right).  
\label{eq:eq381}
\end{equation}
The $\psi$-operators are not eigenvectors of the Hamiltonian and 
thus have no definite frequency.  We can, though, express them 
through the eigenoperators of the Hamiltonian, $\alpha$, 
$\alpha^\dagger$:
\begin{equation}
\alpha_{i,\uparrow}(t)=\alpha_{i,\uparrow}e^{-iE_q t};\; 
\alpha_{i,\downarrow}^\dagger (t)=\alpha_{i,\downarrow}^\dagger 
e^{iE_q t}. 
\label{eq:eq383}
\end{equation}
By doing so, and gathering the terms with $\alpha_{q,\uparrow}$ 
and so on, one obtains the BdG equations 
\begin{equation}
\left( \begin{array}{cc} \hat{t}+Y({\bf r}) & \Delta({\bf r}) \\ 
\Delta^* ({\bf r}) & -\hat{t}-Y({\bf r}) \end{array} \right) 
\left( \begin{array}{c} U_i \\ V_i \end{array} \right) 
=E_i \left( \begin{array}{c} U_i \\ V_i \end{array} \right). 
\label{eq:eq384}
\end{equation}
Solving these equations, one gets both the amplitudes $U$ and $V$ 
along with the excitation energies $E_q$ of the system.  
Once the amplitudes are known, the expectation value of 
any single-particle operator $\cal{O}$ can be obtained from 
\begin{equation}
\langle {\cal{O}} \rangle = 2\sum_{E_i > 0}\int d^3 r 
[U_i^* {\cal{O}}U_i f(E_i) + V_i{\cal{O}}V_i^*(1-f(E_i))], 
\label{eq:eq385}
\end{equation}
where $f(E_i)=(1+\exp(E_i/k_BT)^{-1}$ is the Fermi distribution function, 
$k_B$ is Boltzmann's constant and $T$ is the temperature.  
More explicitly, we write the BdG equations as 
\begin{eqnarray} 
{\biggl ( }{-\hbar^2\nabla^2 \over 2 m } +Y({\bf r}) - E_F {\biggr )} 
U_{i}({\bf r}) +\Delta({\bf r}) V_i({\bf r}) &=& E_i   U_i({\bf r}) 
\label{eq:bog1} \\ 
-{\biggl ( }{-\hbar^2 \nabla^2\over 2 m } +Y({\bf r}) - E_F {\biggr )} 
V_{i}({\bf r}) +\Delta^*({\bf r}) U_i({\bf r}) &=& E_i V_i({\bf r}) 
\label{eq:bog2}  
\end{eqnarray}
where $U$ and $V$ are the quasiparticle amplitudes, 
$E_F$ is the Fermi energy, 
$\Delta({\bf r})$ is the (space-dependent) pairing potential and $m$ is 
the neutron mass. 
In the above equations 
we have replaced the chemical potential with the Fermi energy. The  
resulting violation in particle number conservation is found to be 
very small.  In equations (\ref{eq:bog1},\ref{eq:bog2}) the 
subscript $i$ represents all relevant quantum numbers.   
The pairing potential has to be calculated self-consistently as 
\begin{equation} 
\Delta({\bf r})=|g|\sum_{i;0<|E_i|<\hbar\Omega} U_i({\bf r}) V^*_i({\bf r}) 
(1-2 f(E_i)) 
\label{eq:pairpot} 
\end{equation} 
where the sum is over quasiparticle states 
having energy eigvenvalues with absolute values  
$|E_i|$ smaller than a cutoff $\hbar\Omega$, in our calculations 
taken to be 10 MeV.  
In general the set of quantum numbers $i$ includes the 
principal quantum number, the projection 
of the angular momentum $\mu=\ldots,-3/2,-1/2,1/2,3/2,\ldots$ and the 
wavenumber $k_z$ parallel to the $z$-axis.
As we want to describe neutron vortices with cylindrical symmetry, we 
write the quasiparticle states as 
\begin{eqnarray} 
U_i({\bf r})&=&{1 \over \sqrt{L} }u_{n \mu k_z}(\rho) 
\exp[i(\mu-1/2)\theta] \exp[i(k_z z)]  \label{eq:uampl} \\
V_i({\bf r})&=&{1 \over \sqrt{L} }v_{n \mu k_z}(\rho) 
\exp[i(\mu+1/2)\theta] \exp[i(k_z z)] \label{eq:vampl}  
\end{eqnarray}
where $\rho$, $\theta$,  $z$ are the cylindrical coordinates, 
 $L$ is the length of the cylinder, $n$ is a 
radial quantum number and $\mu$ is half an odd integer.  

The contributions to the pairing potential are calculated in 
each subspace of fixed $\mu$ and $k_z$ 
and are then summed  to give the total pairing potential.  In Paper I 
we considered the simplified case of a pairing potential  
calculated at $k_z=0$. 
This approximation is the one currently  used in self-consistent 
calculations  
of vortices in type-II superconductors, where the electron effective 
mass is anisotropic and increases strongly along directions 
perpendicular to the vortex axis (\cite{gygi}). 
In the case of vortices in neutron matter, variations 
in the effective mass are  expected to be much smaller. 
In the present study we shall therefore 
go beyond this approximation and allow for different values 
of the wavenumber along the vortex axis. 
Since a direct comparison of vortices in neutron matter 
and in superconductors is important, we shall also refer to   
the case with $k_z=0$ whenever such a comparison can be 
revealing.   
 
The angular dependence in equations (\ref{eq:uampl},\ref{eq:vampl})  
of the quasiparticle states follows from  imposing 
a  pairing gap of  the form 
\begin{equation} 
\Delta({\bf r})=\Delta(\rho)\exp[-i\theta] 
\end{equation}    
representing a vortex with one quantum of circulation. 
We are not interested 
in larger winding numbers, since they are expected at much 
higher energies. 
Following (\cite{gygi}) we expand the quasiparticle states in terms of 
cylindrically symmetric Bessel functions and impose the 
boundary condition $\Delta(R)=0$ at the edges 
of a cylinder of radius $R$, where $R$ is  
typically $50$-$100\;{\rm fm}$.  
More specifically, the basis functions are chosen as 
\begin{equation}
\phi_{jm}(\rho)=\frac{\sqrt{2}}{RJ_{m+1}(\alpha_{jm})}J_m\left(
\alpha_{jm}\frac{\rho}{R}\right), 
\;\; j=1,\ldots,N,
\label{eq:basis}
\end{equation}
where $m=\mu\pm \frac{1}{2}$ is an integer and $\alpha_{jm}$ is 
the $j$th zero of the Bessel function $J_m(x)$.  The dimension $N$ 
of the basis is chosen large enough to ensure convergence and 
stability of the quantities of interest.   
The quasiparticle amplitudes are expanded as ($i=\{n \mu k_z\}$)  
\begin{eqnarray}
u_i(\rho)&=&\sum_j c_{ij}\phi_{j\mu-\frac{1}{2}}(\rho) \label{eq:uexp} \\
v_i(\rho)&=&\sum_j d_{ij}\phi_{j\mu+\frac{1}{2}}(\rho). \label{eq:vexp} 
\end{eqnarray}
We consider first the case without mean field, which is the one 
usually addressed in the calculations for type-II superconductors. 
For given values of $\mu$ and $k_z$ equations 
(\ref{eq:bog1},\ref{eq:bog2}) can 
be written as a $2N\times 2N$ matrix eigenvalue problem 
\begin{equation}
\left( \begin{array}{cc} 
       T^- & \Delta \\
       \Delta^T & -T^+ 
 \end{array}\right) \Psi_n=E_n\Psi_n, 
\label{eq:eigval}
\end{equation}
where the superscript $T$ denotes the transpose of a matrix, 
$\Psi_n^T=(c_{n1},\ldots,c_{nN},d_{n1},\ldots,d_{nN})$, 
\begin{equation}
T^{\pm}=\frac{\hbar^2}{2m}\left(\frac{\alpha_{j\mu\pm 1/2}^2}
{R^2}+k_z^2-k_F^2\right)\delta_{jj'},
\label{eq:onebody}
\end{equation}
and the matrix $\Delta$ is given by 
\begin{equation}
\Delta_{jj'}=\int_0^R \phi_{j\mu-1/2}(\rho)\Delta(\rho)\phi_{j'\mu+1/2}
(\rho)\rho d\rho.
\label{eq:pairmat}
\end{equation}  
In cases where a mean field is present the matrix elements in 
equation (\ref{eq:onebody}) get an additional term. 
In our calculations we assume that the mean field of a nucleus  has a 
Woods-Saxon shape, 
\begin{equation}
W(\rho)=\frac{W_0}{1+\exp\left(\frac{\rho-R_N}{a}\right)}
\label{eq:ws}
\end{equation}
where typical values of the constants are 
$W_0\sim -30$ to $\sim - 40\;
{\rm MeV}$, $R_N\sim 5\;{\rm fm}$, and $a\sim 0.6\;{\rm fm}$. 
Note that we only consider cylindrical nuclei in the present paper.  
The case of a spherical nucleus interacting with a cylindrical vortex 
line is quite complicated since the two objects have different 
symmetries.  However, it is worth noting that several calculations 
(\cite{lorenz,petra}) predict that nuclei take on cylindrical shapes 
in parts of the inner crust of a neutron star, so our results 
have in fact a direct bearing on the actual problem. 

Even without a nucleus present, the neutrons move in a 
self-consistent Hartree field of the form 
\begin{equation}
S(\rho)=-|g|n(\rho)
\label{eq:hfield}
\end{equation}
where $n(\rho)$ is the neutron density distribution within the cylinder, 
given by 
\begin{equation}
n({\rho})=2  L^{-1} \sum_{n,\mu,k_z;0<E_{n,\mu,k_z}
<\hbar\Omega}|v_{n,\mu,k_z}({\rho})|^2. 
\label{eq:dens}
\end{equation}
In the presence of a mean field with cylindrical symmetry,  
$ Y(\rho)=W(\rho)+S(\rho) $, the matrix elements of the one-body operator 
in equation (\ref{eq:onebody}) acquire an additional term of the form 
\begin{equation} 
W{^{\pm}_{j j'}}=\mp 
\int_0^R d\rho \rho \phi_{j \mu \pm 1/2} (\rho) Y(\rho) 
\phi_{j' \mu \pm 1/2} (\rho).  
\label{eq:meanmat}
\end{equation} 
For generality 
the expressions have been written at  finite temperature, 
making use of   
the Fermi occupation factors for quasiparticle excitations. 
Although the case at finite temperature presents no 
additional difficulty,   
in applications to neutron star superfluids the temperature 
is essentially zero compared to the Fermi temperature.  Thus, 
all numerical calculations in this paper are carried out at 
$T=0$.  
One can use the symmetries of the BdG equations under the transformation   
\begin{equation} 
(U, V, E) \rightarrow (-V^*, U^*, -E),
\label{eq:transf}
\end{equation} 
i.e., if $(U,V,E)$ is a solution, then so is $(-V^*,U^*,-E)$, 
to rewrite equation (\ref{eq:pairpot}) simply as 
 \begin{equation} 
\Delta({\bf r})=2|g|\sum_{i;0<E_i<\hbar\Omega} U_i({\bf r}) V^*_i({\bf r}) 
=2{|g|\over L} 
e^{-i\theta} \sum_{n,\mu,k_z;0<E_{n,\mu,k_z}<\hbar\Omega} u_{n,\mu,k_z}
({\rho}) v_{n,\mu,k_z}({\rho})
\label{eq:newgap} 
\end{equation} 
where the sum  now extends over positive
eigenstates of the BdG equations only. This symmetry can be in 
principle exploited to 
solve the system ({\ref{eq:eigval}) for only positive 
(or only negative) eigenstates.  

A point-like pairing interaction  
$v({\bf r},{\bf r}')= - |g| \delta({\bf r} - {\bf r}') $ 
has been used in previous work on superconductors (\cite{degennes,gygi,jap}).  
In the case of neutron 
matter this approximation should work whenever 
the range of the inter-particle interaction (which is of 
the order $\sim 1\;{\rm fm}$) is at least 
comparable to or smaller than the nucleus and the vortex.   
Exactly as for a system of electrons, a cutoff energy $\hbar\Omega$ 
has to be introduced for the sum in equation (\ref{eq:newgap})   
to converge. 

\section{Results} 

\subsection{Calculations for different gaps} 

One of the parameters still largely unknown in the physics of neutron 
stars is the value of the neutron pairing gap as a function 
of  the density in the crust and in the interior. The reason 
is that the gap is very sensitive to the value of the 
parameters defining the neutron-neutron interaction at the Fermi surface, 
as can be seen from the weak-coupling formula for the gap 
\begin{equation}
\Delta\approx 2\hbar\Omega\exp\left(-\frac{1}{N(0)\tilde{V}(k_F,k_F)}\right)
\label{eq:wc}
\end{equation} 
where $\hbar\Omega$ is an energy cutoff (often taken to be the Fermi 
energy), $N(0)$ is the density of states for a fixed spin projection at 
the Fermi surface and 
$\tilde{V}$ is the effective interaction between two Cooper-paired neutrons.  
In many studies (Baldo et al. 1990; Elgar\o y \& Hjorth-Jensen 1998;
Khodel, Khodel \& Clark 1996)  
the last quantity is approximated by the free 
neutron-neutron interaction, taken from various nucleon interaction 
models fitted to proton-proton and neutron-proton scattering data.  
The weak-coupling approximation is poor in neutron matter due 
to the strong momentum-dependence of the free neutron-neutron interaction, 
and the BCS gap equation has to be solved in its full 
complexity (\cite{baldo90}).  
While sometimes numerically tricky, this task can now be carried out 
accurately, and the results turn out to be insensitive to the model 
chosen for the nucleon-nucleon interaction as long as it fits the 
scattering data accurately (\cite{khodel96,elg98}).   
However, the interaction between 
two neutrons is modified when other neutrons are present, 
and an important contribution comes from the so-called induced interaction, 
corresponding to the exchange of density and spin-density fluctuations 
between two neutrons (\cite{migdal67}).  
The effect of this has been calculated by Wambach, Ainsworth, \& Pines(1993)   
and \cite{schulze96} with somewhat different results.  
\cite{wambach93}, calculating the induced interaction accurately, 
but using the weak-coupling approximation for the gap,   
predict a decrease by a factor of $\sim 3$ at all densities 
compared with the calculations 
using free nucleon-nucleon interactions.   
\cite{schulze96}, solving the full gap equation, but using a 
somewhat rougher treatment of the induced interaction,  predict a 
similar reduction of the gap at low densities, but an 
increase at higher densities.  
We know of no estimates for contributions beyond the induced interaction; 
some of these may turn out to be important.  Given the uncertainties 
in the predictions for gaps in uniform neutron matter, we will in 
some of our calculations let this be a free parameter and study how 
other quantities depend on the size of the gap.  In other calculations 
we will use the results of \cite{wambach93}, since these are commonly 
employed in other neutron star studies, and have so far been consistent 
with the glitch data. 
Actually, we  find the nice feature that some important properties of the 
vortex are not too sensitive to the gap.  
In a first group of calculations we shall thus keep the value of the 
gap free, using several values of the 
pairing force $|g|$ for a fixed Fermi wave number. 

Keeping comparable values of the gap at infinity and changing $k_F$ 
we find that $|g|$ is approximately inversely proportional to the 
Fermi wave number. This is in rough agreement  
with the  weak coupling formula which predicts (with $\tilde{V}
(k_F,k_F)=-|g|$)  
\begin{equation} 
 |g| k_F = -\frac{2\pi^2\hbar^2}{m}\frac{1}{\ln\left(\frac{\Delta}{
2\hbar\Omega}\right)},   
\end{equation}   
i.e., $|g|k_F\approx {\rm constant}$ modulo a slowly varying logarithmic 
factor.

Figure \ref{fig1} shows the pairing gap as a function of the distance 
from the vortex core. 
As analyzed in Paper I, the gap increases from zero to an asymptotic value 
$\Delta_{\infty}$, which is the value in infinite homogeneous matter.
From the variation of the gap we can extract the appropriate 
size of the vortex. As in Paper I, we take the intersection of the 
tangent line at the origin $\rho=0$ with the line $\Delta_{\infty}$ 
to define the length  $\xi_2$. The quantities 
$\xi_{50}$ and $\xi_{90}$ are the distances  
from the axis where the gap reaches $50\%$ and $90\%$ of the 
value at infinity, respectively. The length $\xi_0$ is simply the 
BCS coherence length, i.e. the size of a Cooper pair.  
In Paper I we showed that with one single $k_z=0$ state 
the size of a vortex tends to change smoothly as a 
function of the pairing force $|g|$ and the Fermi wave number $k_F$.   
Table 1 of the present paper 
collects data for the vortex size for several different Fermi wave 
numbers and asymptotic gaps. 
The data show that the vortex size, although certainly not 
constant, has a limited range 
of variation, while 
 the BCS coherence length changes dramatically. 
For a fixed $k_F$ and increasing 
$\Delta_{\infty}$ the lengths generally tend to decrease.  
The reason for the different scaling of the vortex core compared 
with the BCS coherence length will be examined in Sect. 3.2   
while consequences of 
astrophysical relevance  
will be discussed in Sect. 3.6. 

Figure \ref{fig2} shows the eigenvalues of the vortex state for 
the case of one single 
$k_z$ mode. Similar to that found in other systems 
exhibiting fermionic superfluidity, 
such as superconductors (\cite{ketterson}) or finite 
nuclei (\cite{schuck}), a window 
of width $2\Delta_{\infty}$ opens up where only bound states are present.   
 These  are 
 visible as a branch that  for large angular 
momenta $\mu$ approaches the energy of continuum states,  
very close to  
what one finds for vortices in 
type II superconductors (\cite{gygi}). 
The smallest eigenvalue has 
an energy of the order $\sim \Delta_{\infty}^2/E_F$. For more 
$k_z$ states, as shown in figure \ref{fig3}, there are in general more 
 bound states for a given angular momentum $\mu$  and the energy is roughly  
$E\sim \mu\Delta_{\infty}^2/E_F[1-(k_z/k_F)^2]^{1/2}$ (\cite{caroli}). 
The eigenfunctions of the bound states are confined in the region 
of the vortex core and decrease exponentially far from the core. 
The wavefunctions of the 
continuum  oscillate like scattering states 
with a radial wave number $k_{\rho}$   determined by the condition 
$E_n^2=\Delta_{\infty}^2+(\hbar^2/2m)^2(k_{\rho}^2+k_z^2-k_F^2)^2$.

\subsection{Calculations with fixed gap at infinity} 

In a second group of calculations we assume that the pairing gap 
of neutron matter is given by the results of  (\cite{wambach93}).    
In short,  once the neutron Fermi wavenumber 
is fixed, we choose the value of the 
pairing strength $g$ which reproduces  the given value 
of the pairing gap. This procedure is  necessary  
when a nucleus is present in the center of the vortex core, 
because a nucleus modifies the local density. This in turn 
has strong influence on the local gap and condensation energy 
density, which are both strongly density--dependent.   

We solve the BdG equations for a homogenous 
system at various densities, and require that the energy gaps 
thus obtained be equal to the ones of Wambach et al.  
This provides us with a density dependent pairing strength 
$|g|=|g(n(\rho))|$.  
When solving the BdG equations for a vortex line, we naturally find 
that the neutron density varies within the cylinder.  Thus, $g$ 
will also vary throughout the vortex, and this must be taken into account.  
The pairing potential is then modified to 
\begin{equation}
\Delta({\bf r})=2 \sum_{i;0<E_i<\hbar\Omega}|g({\bf r})|U_i({\bf r})V_i^*
({\bf r}), 
\label{eq:modpairpot}
\end{equation}
and similarly for other quantities involving $|g|$.  

In the numerical solution of the BdG equations, 
we use $50-100$ Gaussian mesh points for the radial coordinate, and  
some $50-100$ angular momentum states and $8-10$ plane wave 
states (for the $z$ direction) in the expansion of the $U$ and 
$V$ amplitudes.  Starting from initial approximations to 
$\Delta(\rho)$ and $n(\rho)$ the BdG equations are solved by 
diagonalizing the resulting eigenvalue problem.  
This gives us approximations to $U$ and $V$, and from these 
new approximations to $\Delta(\rho)$ and $n(\rho)$ are obtained.  
We iterate this procedure until the variations in these quantities 
are small from one iteration to the next.   
Usually this procedure converges after 5-10 iterations, and it 
is adequate to use $\Delta(\rho)={\rm constant}$ and $n(\rho)=
{\rm constant}$ as initial approximations. 
Figures \ref{fig4} and \ref{fig5} show the pairing potential $\Delta(\rho)$ 
and the neutron number density $n(\rho)$ respectively  
as functions of the distance from the axis for 
$k_F=0.8\;{\rm fm}^{-1}$ in four different configurations: an isolated 
vortex at $\rho=0$, an isolated nucleus, vortex and nucleus 
both present, and uniform matter. 
When only the vortex is present 
the gap rises linearly to an asymptotic value within lengths 
specified by the parameters $\xi$ defined in Section 3.1. The fact that   
the pairing strength is now  space-dependent due to 
the decrease  of the density in the core visible in figure \ref{fig5},    
changes only slightly the  values of the vortex core radius.    
The shape of the gap is also only slightly altered by the 
variation of the pairing strength. The case where only the mean field 
is present shows a decrease of the pairing gap in the 
region occupied by the nucleus, 
qualitatively consistent 
with semiclassical approximations (\cite{semiclassical}).   
When both a nucleus and vortex are present the picture becomes more complex 
due to the strong distortion of the  velocity field induced 
by the nucleus.  To understand the difference in behavior of the gap  in 
the presence of a vortex,  with or without the nuclear mean field, we 
analyze the problem using the Ginzburg-Landau (GL) equation     
\begin{equation}
-\frac{\hbar^2 \nabla^2}{4m}\psi({\bf r})+W({\bf r})\psi({\bf r}) 
-A\psi({\bf r})-B|\psi({\bf r})|^2\psi({\bf r})=0
\label{eq:gl1}
\end{equation} 
where $\psi({\bf r})$ is the order parameter 
and the self-consistent mean field $S(\rho)$ is neglected compared 
to $W(\rho)$, which is a very reasonable approximation.  
The parameter $A$ is negative for $T<T_C$, where $T_C$ is the 
critical temperature of the superfluid, while $B$ is always positive. 
Using cylindrical coordinates, taking $\psi({\bf r})=e^{-i\theta}f(\rho)$, 
$W({\bf r})=W(\rho)$, and using units where $\hbar ^2/4m=1$, 
equation (\ref{eq:gl1}) can be 
written as 
\begin{equation}
f''(\rho)+\frac{1}{\rho}f'(\rho)+
\left[ -W(\rho)-A-B|f(\rho)|^2-\frac{1}{\rho^2} \right]f(\rho)=0.
\label{eq:gl2}
\end{equation}
The boundary condition at $\rho=0$ is $f(0)=0$.  Looking at small $\rho$, 
we neglect the non-linear term, and distinguish between two cases.  
First, for the case of no nuclear mean field, equation (\ref{eq:gl2}) 
becomes 
\begin{equation}
f''(\rho)+\frac{1}{\rho}f'(\rho)-\frac{1}{\rho^2}f(\rho)=0,
\label{eq:gl3}
\end{equation}
since $A$ is supposed to be small near $T_C$ and thus negligible compared 
to $1/\rho^2$ for small $\rho$.
This equation is easily seen to have the solution $f(\rho)\propto \rho$.  
Thus, we expect the pairing potential to show a 
linear increase for small $\rho$ in the case of an isolated vortex, 
and this behaviour is also seen in our microscopic calculations.  
Looking at the second case, an attractive mean field $W(\rho)
\approx -W_0$, assumed to be constant for the limited range in $\rho$ 
we are looking at now, equation (\ref{eq:gl2}) is reduced to 
\begin{equation}
f''(\rho)+\frac{1}{\rho}f'(\rho)+\left(W_0-\frac{1}{\rho^2}\right)
f(\rho)=0, 
\label{eq:gl4}
\end{equation}
where we have absorbed the constant $A$ in the mean field $W_0$.  
Introducing a new variable through $\rho=\frac{1}{\sqrt{W_0}}x$, 
we obtain 
\begin{equation}
x^2f''(x)+xf'(x)+(x^2-1)f(x)=0,
\label{eq:gl5}
\end{equation}
which the reader may recognize as the Bessel equation of order 1.  
The solution satisfying $f(0)=0$ is $f(x)\propto J_1(x)$.  
Thus, in the case of a vortex interacting with a nucleus, we 
may expect the pairing potential to show oscillations at small $\rho$.  

Figure \ref{fig6} shows the relevant lengths for the case of 
polarized gap. The lengths remain limited ($<15$ fm) for 
essentially all densities and tend to decrease at higher 
density. This behavior is different from the 
BCS coherence length $\xi_0$ which scales like  $k_F/\Delta_{\infty}$.  
In the realistic case where the pairing strength is fitted to the 
gap in uniform matter, the coherence length $\xi_2$ decreases as a 
function of $k_F$.   
Thus, the density dependence of the pairing strength $|g|$  
is also very important, and  
the vortex size appears to depend in a quite complex way on both $|g|$ 
and the Fermi wavenumber. In the figure we also show  
the analytic estimate of the vortex core size 
made  by \cite{kramer} whose value in our units (energies in 
MeV and lengths in fm) is 
\begin{equation}
\xi_1={661.3\over k_F^2 |g|}.  
\end{equation}

Since  the pinning force between  
a nucleus and a vortex behaves like 
$ \sim \Delta^2 k_F/ \xi_2 $, an increase of 
the pinning force of the order $\sim \xi_0/\xi_2$ can 
be expected. In addition we can confirm what we found in Paper I, that 
since the diameter of the vortex  is always smaller than the lattice 
spacing, 
the vortex envelops at most one nucleus along a plane perpendicular 
to its axis.   

Another interesting parameter  is represented by the coherent flux 
of  neutrons in the presence of a vortex. 
The flux is given by 
\begin{equation} 
J({\bf r})={\hbar \over 2 m i}{\biggl (}
\sum_{i}{\biggl [}V_i({\bf r})\nabla V^*_i({\bf r})-
 V^*_i({\bf r})\nabla V_i({\bf r}){\biggr ]}(1-f(E_i)) 
+
{\biggl [}U_i({\bf r})\nabla U^*_i({\bf r})-
 U^*_i({\bf r})\nabla U_i({\bf r}){\biggr ]}(1-f(E_i)) 
{\biggr )}
\label{eq:flux1} 
\end{equation} 

which in the present case   becomes 

\begin{equation}
J(\rho)={\hbar \over 2 m \rho} {\biggl [}
\sum_{E_i<0} (\mu - 1/2) (|u_i({\rho })|^2 
-\sum_{E_i>0} (\mu + 1/2) |v_i({\rho })|^2 
{\biggr ]}. 
\end{equation}

Figure \ref{fig7} shows the flux of neutrons with and without 
a nucleus present in the middle of the cell for the case 
of $k_F=0.4$ fm$^{-1}$. The main effect of the nucleus is to decrease 
the  flux. The coherent velocity field can be obtained 
dividing the flux by the neutron density. In this case the 
velocity is even more depleted in the centre due to the much higher 
density in the nuclear region than outside, a  situation typical  
for low values of $k_F$.   

\subsection{Variations in $\Delta$ due to a center of scattering} 

Pinning or antipinning between a vortex and a nucleus is due  
to local variations in both the kinetic energy density 
and the condensation energy. Although pinning will be 
examined in detail in a next section, it is interesting 
to study how the pairing gap changes in the cell when a nucleus 
alone is present. Some curves have already been presented in Figure 1, but 
here we shall be more systematic discussing of the applications to 
pinning calculations.  The variation of the pairing gap as 
a function of the distance from the axis 
of a cylindrical nucleus 
when the vortex is absent 
can be seen in Figure \ref{fig8} for some selected densities. Note 
that the presence of a mean field decreases the value of the 
gap. This is mainly due to the increase of the density 
in the nuclear region, since  the pairing force $|g|$ depends on the 
local density. Note also that the variation of the 
gap extends  beyond the range of the mean field, indicated with a 
dashed line. 
This is due in part to the long tail   
of the density distribution beyond the  radius of the potential,  
an effect due to 
 the presence of scattering states. Secondly, there is a proximity effect 
between the nuclear region and the free neutron gas due to  
Andreev scattering (that is, scattering due to spatial 
variations  the pairing potential) of quasi-particles.

\subsection{Pinning energy and vortex self-energy}

The vortex self-energy or  
tension $T$, which is the energy carried by the vortex 
per unit length, is a relevant parameter for vortex-nucleus pinning.  
Let us consider the behavior of a single 
vortex line in interaction with the whole lattice.  
If the energy necessary for the vortex to reach the pinning centers is 
very high compared to the energy gained  by pinning, the vortex 
 responds   stiffly to local deformations. The importance of this 
parameter can be grasped in a limiting case: for $T\rightarrow \infty$ the 
vortex behaves like a rigid cylinder. For a vortex 
moving perpendicularly to its axis  the 
total pinning force with the lattice arises  
from stochastic summation from all the pinning centers 
and  is proportional to the square root of the vortex 
length (see for example \cite{tinkham} for 
the case of superconductors). 
Thus, the pinning force  cannot  balance  
  the Magnus force, which increases 
linearly with the vortex length. If this was the case,  
the  pinning mechanism would not be effective for storing 
the energy released during a glitch  (\cite{anderson}). 
For a finite stiffness, the situation is more complicated.  
Solving the  
equation of motion for a vortex line passing through many centers of pinning 
would give the value of  the total pinning 
energy which is able to be stored, but this 
is a difficult task due to the difficulty in finding the  appropriate 
boundary conditions in the presence of many centers of pinning.     
 It has been argued that although finite, 
the tension is too large to account for large glitches 
(\cite{jones98}). If, on the other hand, the pinning model is   
correct, the value of the vortex tension is relevant    
for the dynamics of unpinning (\cite{link}) and for the 
value of the effective pinning force per unit length.

The vortex tension is 
the energy difference per unit length between the configuration 
with and  without a vortex,  
$
T=\left [E({\rm vortex})-E({\rm uniform\; matter})\right]/L.    
$
To calculate the energy of the superfluid  in each 
configuration one can take  
averages of Eq. (6). The result is 
\begin{equation}
E=\int d^3 r {|\Delta({\bf r})|^2\over |g|}
+2\sum_{E_n<0} E_n-2\sum_{E_n}E_n\int d^3 r |V_n({\bf r})|^2.  
\label{energy}
\end{equation}
Table (2) shows the calculated points as a function of the Fermi 
wave number compared to the classical calculation where 
the kinetic energy density is integrated from the coherence 
length $\xi$ up to  an upper limit $R_0$: 
\begin{equation} 
T=4.38 k_F^3 \ln{\biggl (}{R_0\over \xi}{\biggr )}. 
\label{eq:testimate} 
\end{equation}
Evidently the presence of bound states does not change the 
vortex tension radically. Furthermore, constraining the 
vortex in a small cylinder can overestimate the role of bound 
states in the calculation of self-energy, because 
 the vortex is macroscopic 
along the $\rho$ direction. 
We also expect that uncertainties in the 
pairing gap or in the cutoff length $R_0$ (that should be of the 
order of the vortex-vortex separation) may be much greater than the 
discrepancy we find between our microscopic calculations 
and the estimate (\ref{eq:testimate}). We conclude  that 
equation (\ref{eq:testimate}) is reliable when applied to 
vortex calculations in a neutron star. Assuming a vortex-vortex 
separation of the order $R_0\sim 10^{-2}$ cm, this formula implies 
a tension of the order $T\sim 10^9$ erg cm$^{-1}$ at $k_F=0.8$ fm$^{-1}$. 
At such values a vortex can deform only at length scales of the 
order of several hundreds of fm, nearly $20$ times larger 
than the lattice spacing, unless the pinning energy assumes 
unrealistically large values. 
How a vortex line with such small deformability 
can adapt its shape to gain a net pinning energy 
 is still an unsolved problem  
(\cite{jones97}).  
We show in the following  that an answer may be 
provided by the high value of the elementary pinning force. 

To calculate the pinning energy we need to consider 
four different configurations: a vortex superimposed on  
a nucleus, an isolated nucleus, an isolated vortex, and finally 
uniform matter. For each configuration, the energy needs to 
be calculated with Eq.  (\ref{energy}). 
The pinning energy per unit length can be calculated as 
\begin{equation}
E_p={\biggl [}E({\rm nucleus\;and\;vortex})-
E({\rm nucleus})+E({\rm uniform\; matter})-E({\rm vortex})
{\biggr ]}/L. 
\end{equation}

Table (3) collects some pinning data calculated with Eq. (49) 
and (51). A negative sign corresponds to a pinning 
situation in which it is energetically favorable for a vortex to 
remain attached to a nucleus, while a positive sign 
indicates an {\it antipinning} situation, where vortices 
are repelled by nuclei. We find an antipinning-pinning transition 
 at $k_F=0.30-0.40$ fm$^{-1}$. 
A similar transition was found in a Ginzburg-Landau approach     
 (\cite{baym}). The  pinning energy 
reaches values  of 
the order of several MeV per unit length,  corresponding 
to a very high value if  
   compared 
to  previous calculations performed solely on the 
basis of the condensation energy loss. As is also found in 
other models where the kinetic energy is  accounted for 
(\cite{baym}), part of these large values 
can be attributed to kinetic energy effects.  
For such large values of the pinning energy a vortex is likely to 
deform sensitively in order to be able to catch the pinning centers. In addition, 
the ions in the lattice will shift to the vortex in response to the 
high pinning force, a situation referred to as superstrong pinning. 
Since  nuclei in most of the crust are spherical, one might 
question  the applicability of our calculations, where 
 nuclei are considered to be cylindrical. Although the proper 
use of spherical geometry would, of course,  quantitatively change 
the results, we do not expect dramatic differences.

\section{Conclusions} 

The results presented 
in this paper could be a step towards 
an understanding of vortex structure and interaction 
in neutron star crusts. We have  
found  differences 
between the predictions of microscopic theory and those of more 
macroscopic models.   In particular, we showed that the 
BCS coherence length cannot be used as a reliable estimate of the 
vortex core size, especially not at low and 
intermediate densities.  By solving the self-consistent 
problem of a vortex in neutron matter, we have shown 
that the vortex core has a 
much more complicated structure than that which is usually assumed in 
simplified treatments of the problem, e.g.: 1) the gap changes 
from zero at the axis of the core up to an asymptotic value, 
and 2): there are bound states present.   

We found that vortices in neutron star matter follow a behavior similar 
to that described by \cite{kramer} for superconductors, 
namely, at zero temperature the vortex core size scales as the 
inverse of the Fermi wave number. In a superconductor 
in clean limit this effect is associated with a linear increase of the 
vortex core size with temperature. We have not investigated the temperature 
dependence of our results, because this is not very 
relevant in neutron star physics due to the low temperatures 
compared with the Fermi energy. Unfortunately, experimental data for 
clean type II superconductors  are limited to the  
quite complex new cuprate systems. Scanning tunnelling microscopy has, 
however, confirmed the presence of bound states of quasiparticles in the vortex core, in 
 good agreement with the Bogoljubov-de Gennes theory.  In view of 
the small range of the nuclear force and of the simple structure of the Fermi 
surface, the BdG model could be a very good approximation for 
neutron star matter. 
 
High pinning energies and small lengths imply a large 
pinning (or anti-pinning) force. In addition, the small value of 
the coherence length  means that a vortex envelopes no more than one 
 nucleus in a plane perpendicular to the vortex axis. 
The possibility for a vortex to store pinning energy is 
largely the result of a competition between the energy gain with pinning 
and the energy cost to deform. 
To be able to deform and catch 
a nucleus along its path, a vortex has to be relatively free 
to perform local deformations. 
Assuming that the energy of the vortex 
is solely due to kinetic effects, the 
tension turns out to be of the order $T\approx 100 k_F^3$ MeV fm$^{-1}$  
and is thus very high  
 at densities corresponding to the deep inner crust. 
For a total pinning energy per nucleus $E_p$, a vortex can 
efficiently catch the pinning sites only if they are 
at least a distance $\approx a^2 T/2$ apart. 
  For example, at $k_F=1$ fm$^{-1}$ a vortex deformation 
of the order of the lattice spacing can occur every $\sim 8$ 
or so lattice spacings, while at $k_F=0.8$ fm$^{-1}$ the 
pinning efficiency would be much better.  
Calculations of the geometry of vortex 
deformation  in a one-dimensional crystalline 
structure have been performed by (\cite{link}). 

Finally, there are other unsolved problems in neutron star dynamics where 
a microscopic description of vortex states might prove 
helpful.  
The scattering of electrons off  neutron vortices 
results in a coupling between protons and neutrons in the 
interior of the star. 
 This is a possible  
  coupling mode between the superfluid 
neutrons and the rest of the star,  and depends on the microscopic 
structure of vortex lines (\cite{sauls}). 

The proton vortices 
in the interior of the star induced by the rotation  
will probably be similar to the neutron vortices   
examined in this paper, 
but additionally they will generate a  magnetic field 
parallel to the rotational axis. How these magnetic 
vortices interact with the (much more numerous) flux tubes induced by the 
high magnetic field penetrating into the star is only one 
of the  problems connected to the dynamics of 
superfluid vortices  below the star's crust.  
Despite their importance in the evolution of the 
pulsar's magnetic field, 
the properties of proton flux tubes and their 
possible 
interaction with neutron vortices in the interior, where neutrons 
couple in a triplet phase, are 
not well understood (\cite{ruderman}). A microscopic 
description of proton flux lines might also be applied to the 
study uf ultra-high magnetic fields and the evolution of 
magnetic field in  magnetars.

\begin{acknowledgements}
We would like to thank M. Hjorth Jensen, E. Osnes, 
L. Engvik and T. Engeland 
for illuminating discussions on the neutron star 
equation of state and on   superfluidity 
in both neutron stars and  finite nuclei,  
A. Sedrakian and C. Pethick for  useful  
comments on the present model and M. Colpi for 
an interesting discussion about strong magnetic fields 
in collapsed stars. 
F. V. De Blasio was supported by the Marie Curie 
Research Training Grant under Contract No. ERB4001GT96383G. 
\end{acknowledgements}


\clearpage


\begin{table*}
\caption[]{
Gap at infinity $\Delta_{\infty}$ and the length scales  
$\xi_0$,  $\xi_2$, $\xi_{90}$ and $2\xi_{50}$  for different values of 
the Fermi wave number and pairing gaps.}
\begin{flushleft}
\begin{tabular}{llllll}
\hline\noalign{\smallskip}
{$k_F\;({\rm fm}^{-1}$)}&
{$\Delta_\infty$}&
{$\xi_0$ (fm)}&
{$\xi_2$ (fm)}&
{$\xi_{90}$ (fm)} &
{2$\xi_{50}$ (fm)}
\\ 
\noalign{\smallskip}
\hline\noalign{\smallskip}
   0.1   & 0.0049  & 344.9 & 6.02  & 11.11 & 7.15   \\   
   0.1   & 0.011  & 153.4  & 6.07   & 19.89 & 12.29   \\
   0.2   & 0.02  & 152.0  & 9.80  & 15.61 & 11.03   \\ 
   0.2   & 0.04   & 81.83  &  12.73  &  16.56  &  13.20 \\
   0.2   & 0.183  &  18.41  & 5.57 & 8.04 & 6.04  \\
   0.2   & 0.291  & 11.57 &  5.20 &  7.13 &  5.63   \\
   0.2  &  0.744 & 4.54 & 4.59 & 6.18 & 4.96 \\
   0.5  & 0.226 & 37.34 & 5.32  & 17.71 & 7.266 \\
   0.5  &  0.778 & 10.71 & 2.96 & 9.13 & 3.33 \\
   0.5  & 1.73 & 4.87 & 2.52 & 6.75 & 2.67 \\
   0.5  & 3.42 & 2.46 & 2.09 & 3.24 & 3.02 \\
   0.8  & 0.205 &  65.8 & 6.05 & 24.60 & 17.48 \\
   0.8  & 1.72 & 7.84 & 2.21 & 9.21 & 2.65  \\
   0.8  & 3.8 & 3.55 & 1.67 & 5.02 & 1.87  \\   
\noalign{\smallskip} 
\hline 
\end{tabular}
\end{flushleft}
\end{table*} 

\begin{table}
\caption[]{
Vortex self-energy $T$ (in MeV fm$^{-1}$) at different Fermi 
wave numbers calculated from the program and the value $T_{f}$ 
calculated from the analytical formula, where 
the  coherence length $\xi_2$ has been used 
for the lower length $b$. 
 The radius $R$ of the cylinder  
 is reported.  
\label{tab2}}
\begin{flushleft}
\begin{tabular}{llllll}
\hline\noalign{\smallskip}
{$k_F\;({\rm fm}^{-1}$)}&
{$T$ (MeV fm $^{-1}$)}&
{$T_f$ (MeV fm$^{-1}$) }&
{$\xi_0$ (fm)}
\\ 
\hline\noalign{\smallskip}
   0.4   & 1.08  & 0.64 & 80      \\   
   0.5 & 1.03 & 1.344 & 80 \\
   0.7 & 1.67 & 5.30 & 80 \\
   0.85 & 3.06 & 7.401 & 60 \\
\noalign{\smallskip}
\hline 
\end{tabular}
\end{flushleft}
\end{table}

\begin{table}
\caption[]{
The  pinning energy  per unit length 
$E_p$ (in MeV fm$^{-1}$) at different Fermi 
wave numbers calculated  for a potential well 
with the following parameters: radius $R_N=5$ fm, depth 
$W=-30$ MeV and diffusivity $a=0.6$ fm.  
 The radius $R$ of the cylinder  
 is $80$ fm.  
\label{tab3}}
\begin{flushleft}
\begin{tabular}{llllll}
\hline\noalign{\smallskip}
{$k_F\;({\rm fm}^{-1}$)}&
{$E_p$ (MeV fm$^{-1}$)}
\\ 
   0.15 &   0.68 \\
   0.20 &   0.201 \\
   0.30 &   -0.091 \\
   0.5 & -0.77 \\
   0.7 & -2.34 \\
   0.85 & -16.47 \\
   1.00 & -4.09 \\ 
\noalign{\smallskip}
\hline 
\end{tabular}
\end{flushleft}
\end{table}

\begin{figure}
  \resizebox{\hsize}{!}{\includegraphics{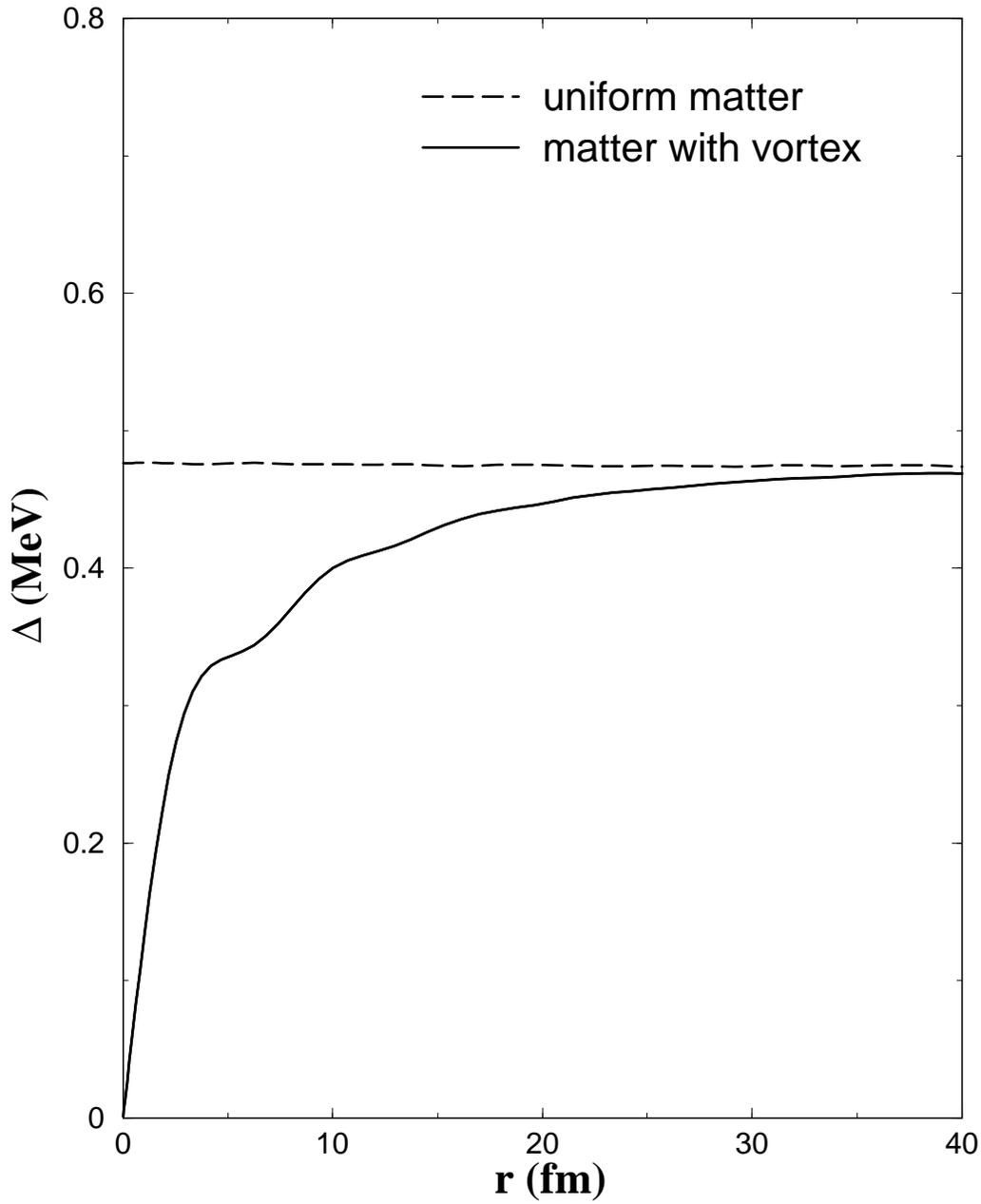}}
  \caption{Pairing potential $\Delta(\rho)$ as a function of the 
           distance $\rho$ from the symmetry axis of the vortex 
           with a pairing force independent of the density and fixed 
     	   to return the desired value of the gap at infinity.}
  \label{fig1}
\end{figure}

\begin{figure}
  \resizebox{\hsize}{!}{\includegraphics{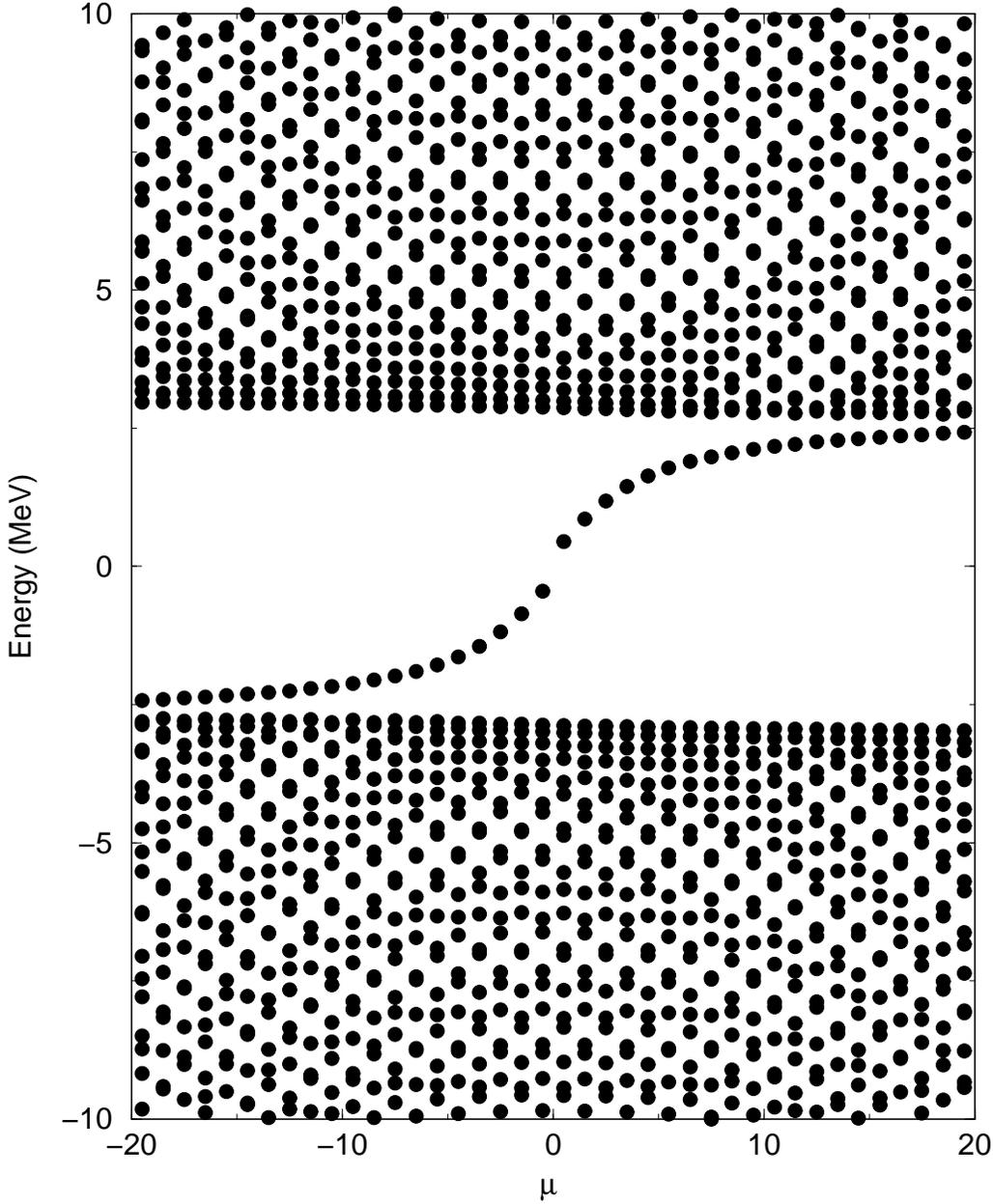}}
  \caption{
       The eigenvalues of the BdG equations for a 
       vortex in neutron matter for  the 
       case $k_F=0.8$ and a pairing gap at infinity 
       $\Delta_{\infty}$=2.9 MeV. The  positive and negative quasiparticle 
       states in the continuum are separated  
       by a gap equal to $2\Delta_{\infty}$. The 
       bound states are visible as the branch between the continuum states.
       Only one single value of 
       $k_z=0$ has been included in the calculation.}
  \label{fig2}
\end{figure}

\begin{figure}
  \resizebox{\hsize}{!}{\includegraphics{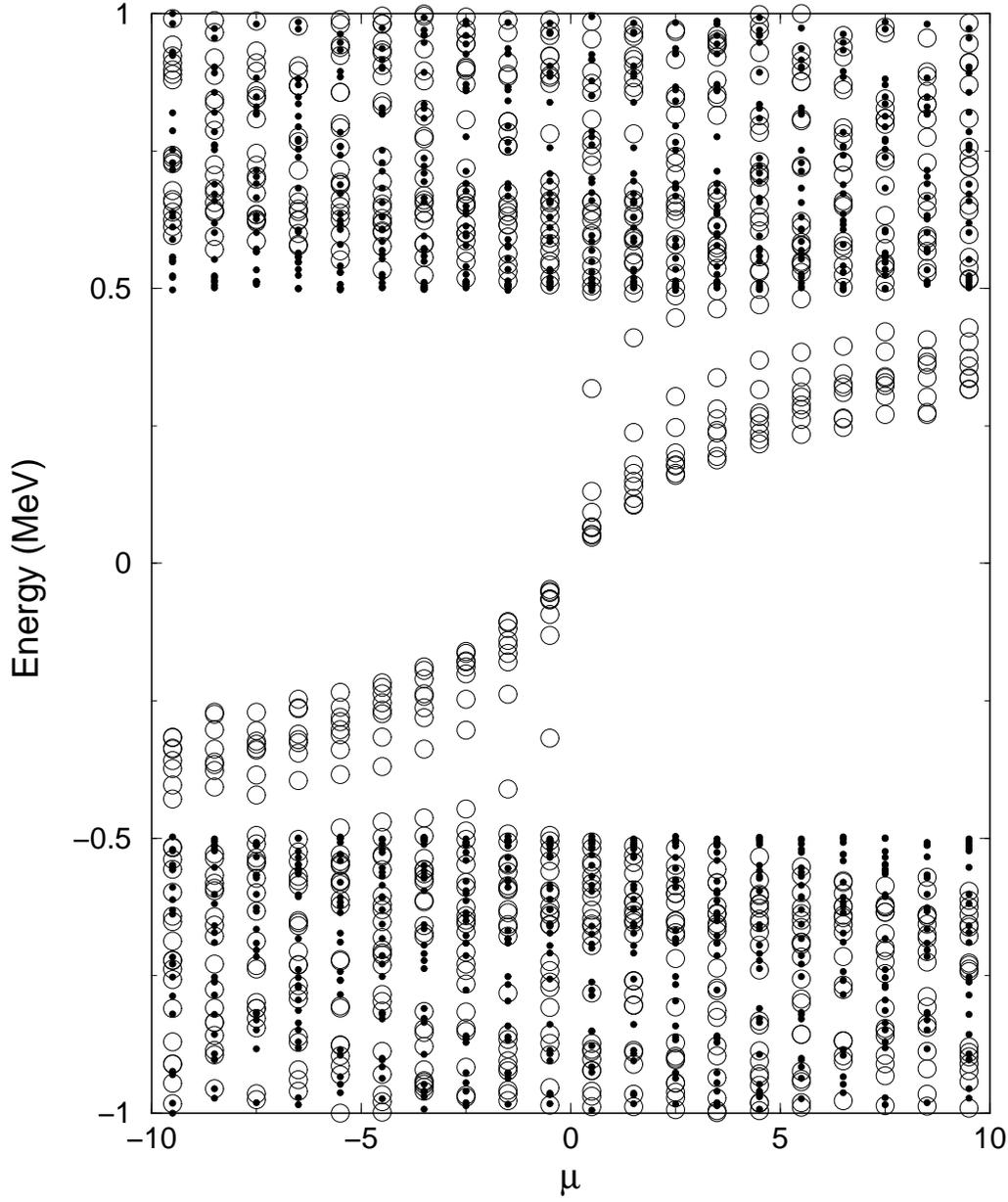}}
   \caption{
     Distribution of energy eigenvalues 
     as a function of angular momentum, now also with different values 
     for $k_z$. $\Delta_{\infty}=0.5$ MeV and $k_F=0.22$. 
    }
  \label{fig3}
\end{figure}

\begin{figure}
  \resizebox{\hsize}{!}{\includegraphics{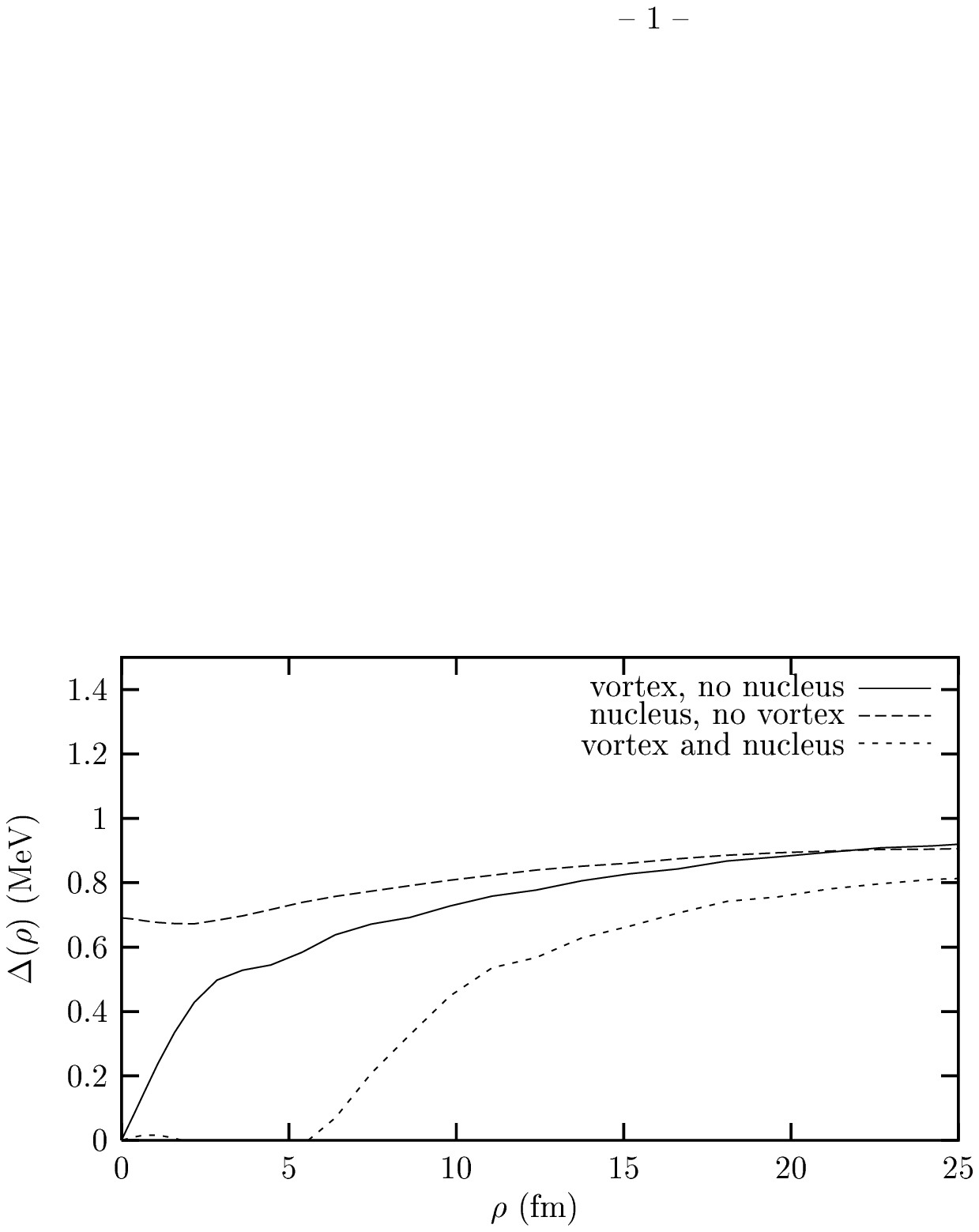}}
  \caption{
    Pairing potential $\Delta(\rho)$ as 
    a function of $\rho$ for three different situations.  The density 
    of uniform matter is in all three cases given by $k_F=0.8\;{\rm fm}
    ^{-1}$.   
   }
  \label{fig4}
\end{figure}

\begin{figure}
  \resizebox{\hsize}{!}{\includegraphics{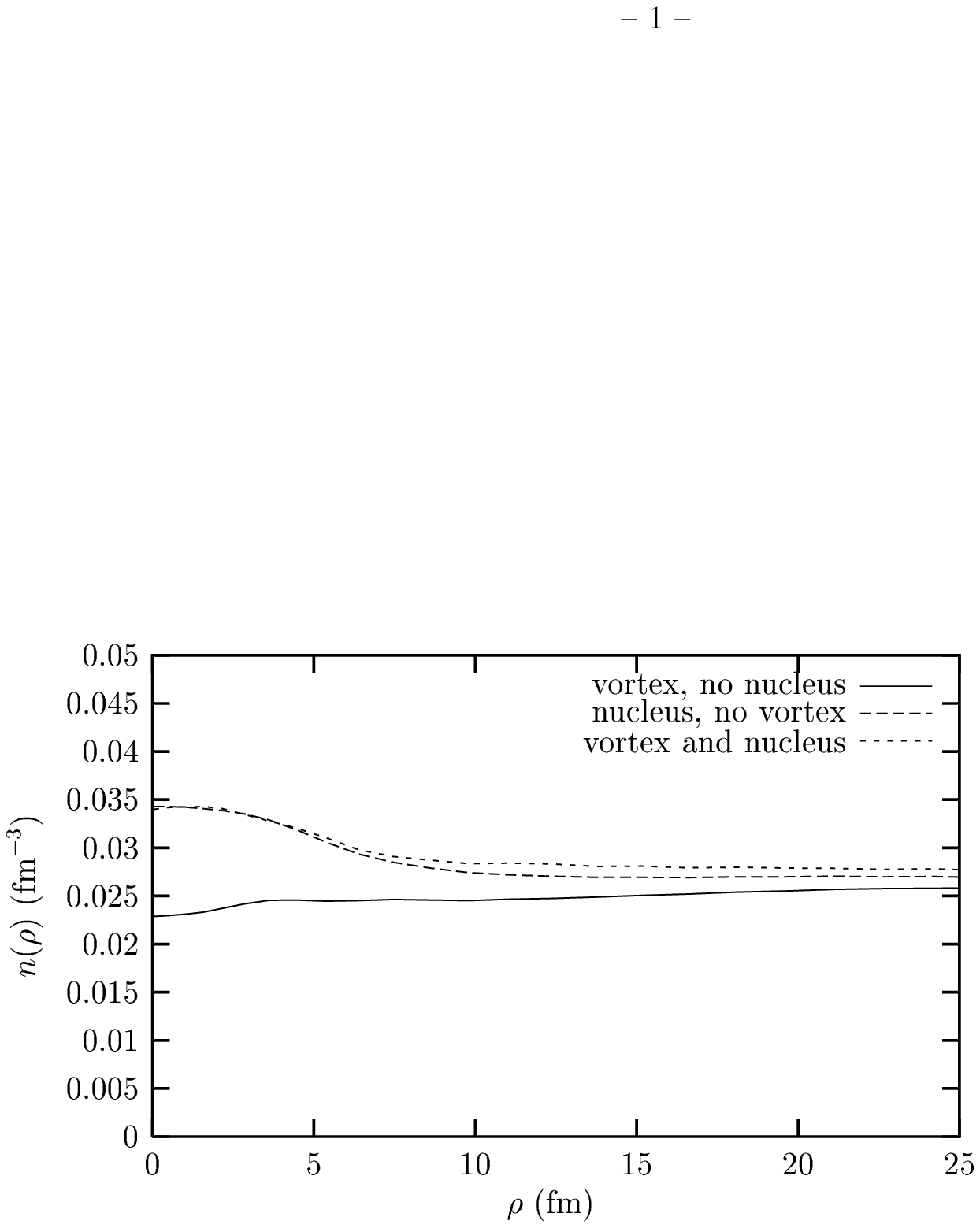}}
  \caption{
    Density distribution $n(\rho)$ as a 
    function of $\rho$ for three different situations.  The density 
    of uniform matter is in all cases given by $k_F=0.8\;{\rm fm}
    ^{-1}$.
   }
  \label{fig5}
\end{figure}

\begin{figure}
  \resizebox{\hsize}{!}{\includegraphics{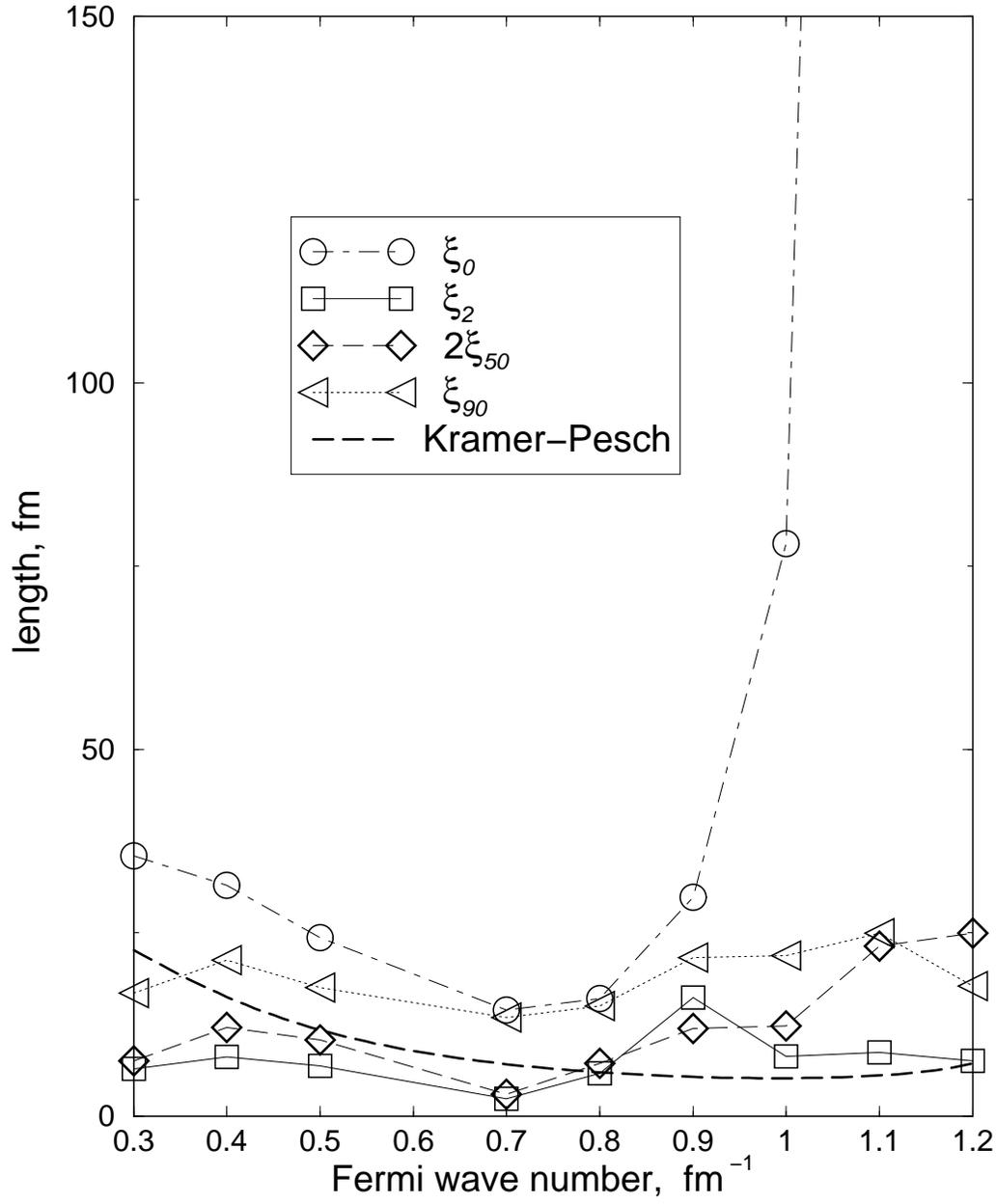}}
  \caption{
    Various length scales for a vortex 
    as a function of the density.  The meaning of the different 
    symbols are explained in the text.
   }
  \label{fig6}
\end{figure}

\begin{figure}
  \resizebox{\hsize}{!}{\includegraphics{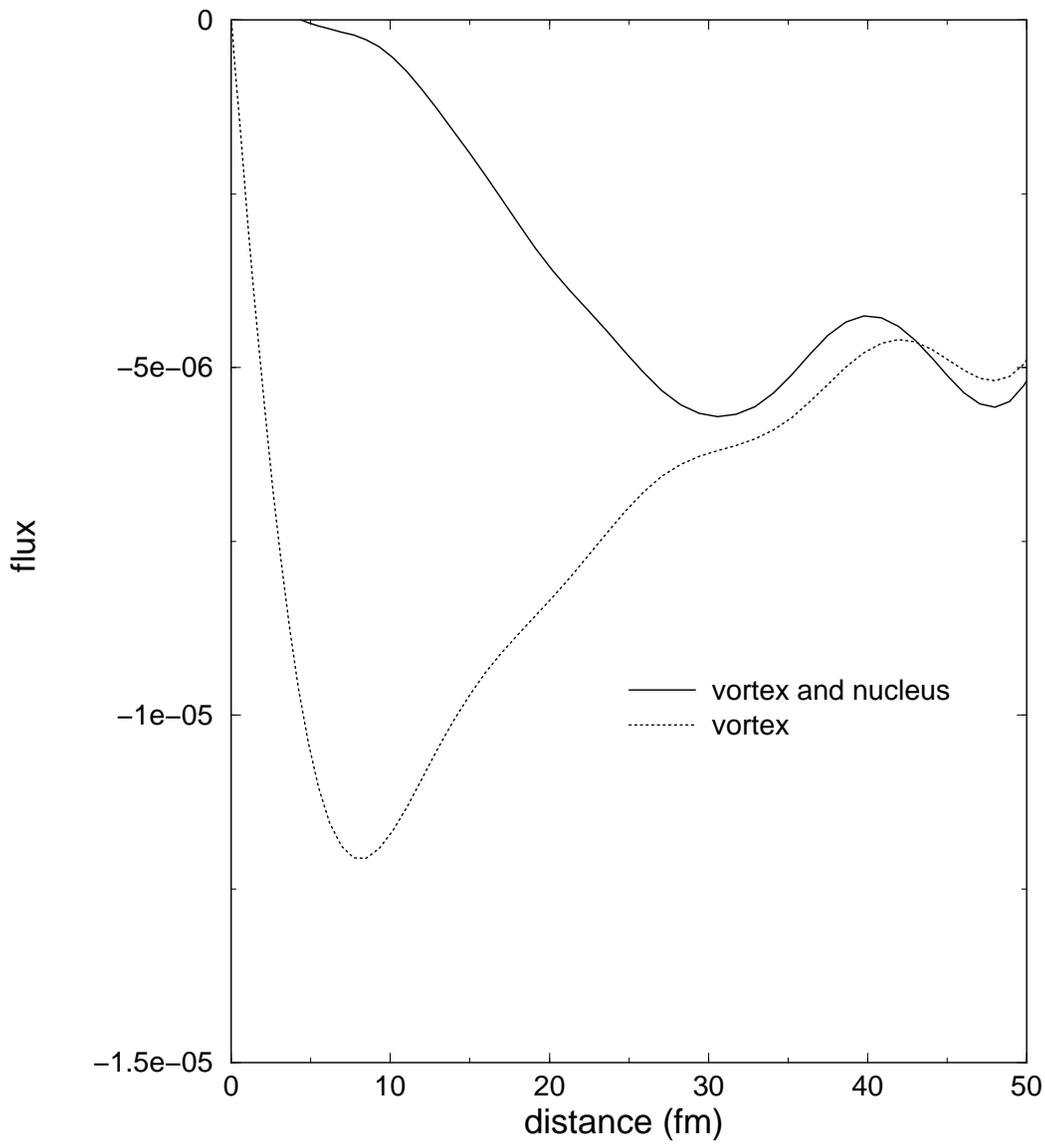}}
  \caption{
    The coherent velocity field for a vortex  with  
    and without a center of scattering. 
   } 
  \label{fig7}
\end{figure}

\begin{figure}
  \resizebox{\hsize}{!}{\includegraphics{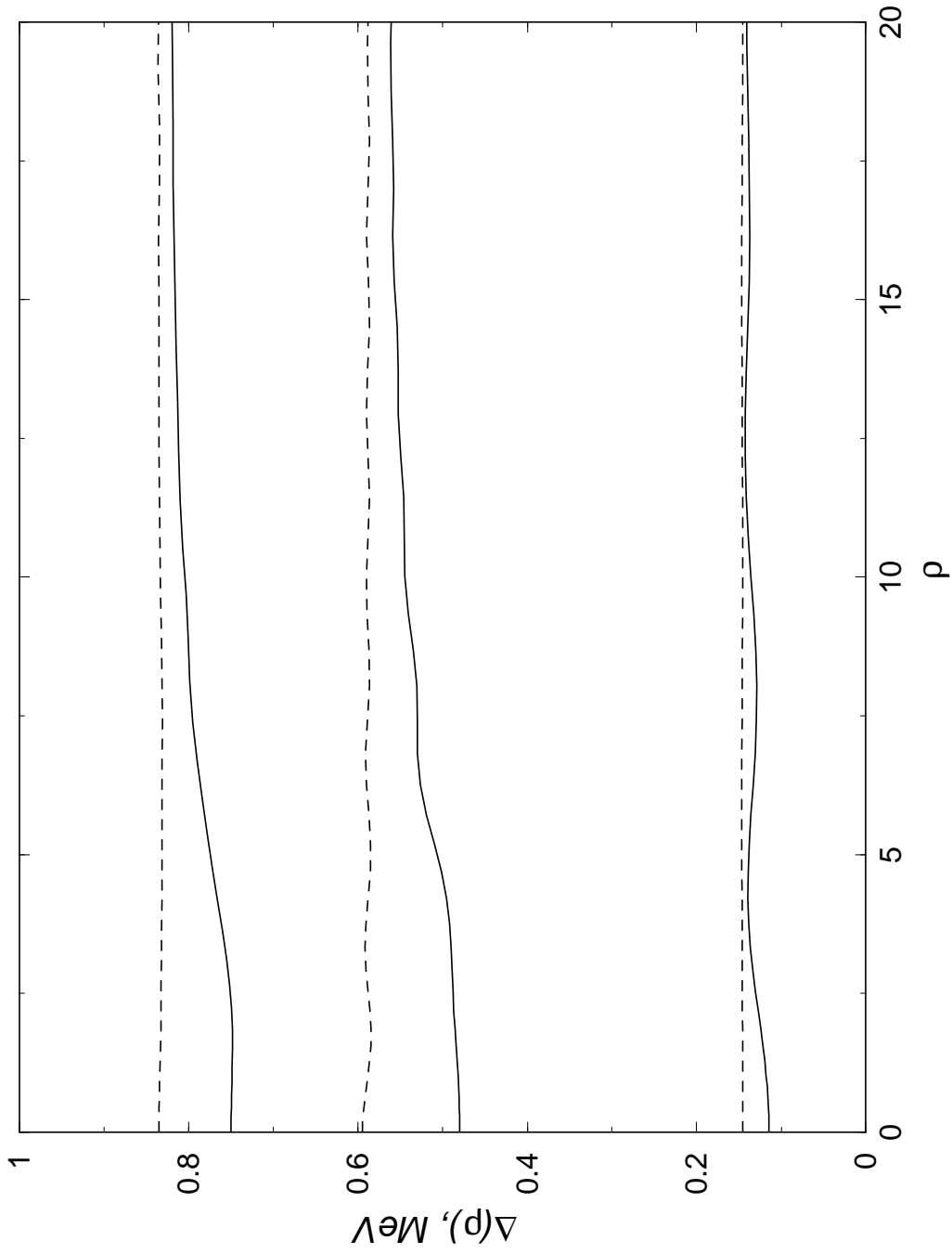}}
  \caption{
   Pairing potential at various 
   densities for the case of a nucleus immersed in a homogeneous 
   (vortex-free) neutron superfluid.
   }
  \label{fig8}
\end{figure}

\end{document}